**Title**

CNFET-based design of efficient ternary half adder and 1-trit multiplier circuits using dynamic logic

**Authors**

Farzin Mahboob Sardroudi[a] – Mehdi Habibi[b] - Mohammad Hossein Moaiyeri[c]

[a] Department of Electrical Engineering, University of Isfahan, Isfahan, Iran

(f.mahboob@eng.ui.ac.ir)

[b] Department of Electrical Engineering, Sensors and Interfaces Research Group, University of Isfahan, Isfahan, Iran (mhabibi@eng.ui.ac.ir) (Corresponding Author)

[c] Faculty of Electrical Engineering, Shahid Beheshti University, Tehran, Iran

(h_moaiyeri@sbu.ac.ir)

**Abstract**

This paper presents a ternary half adder and a 1-trit multiplier using carbon nanotube transistors. The proposed circuits are designed using pass transistor logic and dynamic logic. Ternary logic uses less connections than binary logic, and less voltage changes are required for the same amount of data transmission. Carbon nanotube transistors have advantages over MOSFETs, such as the same mobility for electrons and holes, the ability to adjust the threshold voltage by changing the nanotube diameter, and less leakage power. The proposed half adder has lower power consumption, delay, and fewer transistors compared to recent ternary half adders that use similar design methods. The proposed 1-trit multiplier also has a lower delay than other designs. Moreover, these advantages are achieved over a wide supply voltage range, operating temperatures, and output loads. The design is also more robust to process variations than the nearest design in terms of PDP.

**1. Introduction**

Consecutive challenges of reducing MOS transistor's size, such as short channel effect, high leakage power, reduced gate control, and parameter variations, have forced designers to replace MOSFETs with new alternative technologies [1]. Graphene Nano-Ribbon Field-Effect Transistors (GNRFET) and Carbon Nanotube Field-Effect Transistors (CNFET) are two of these emerging nanotechnologies [2-3]. CNFET seems to be the most viable candidate due to its unique features and functional similarities to MOS transistors. In addition, the one-dimensional band structure and near-ballistic performance of CNFETs improve the performance and energy efficiency of CNFET-based circuits compared to MOSFET-based designs. Therefore, CNFET can be used effectively to design integrated circuits with high energy efficiency [4,5].

In addition to using new low-power devices such as CNFET, another way to reduce energy consumption is to use multi-valued logic (using more than two logic levels to perform calculations). Using multi-valued logic (MVL) design reduces the area, the number of pins, and connections by carrying more information in MVL logic compared with binary logic [6]. Studies have shown that using the base e (approximately 2.718) for calculations results in the highest energy efficiency at the system level [7]. However, due to hardware limitations, the basis of mathematical operations must have an integer value, and the closest integer to the optimal basis can be used. As a result, base 3 operations seem more appropriate for less complexity [8]. The main problem with MVL is less noise margin than with binary logic [9].

Half adders and 1-trit multipliers are used in larger structures such as arithmetic processing units and play an essential role in VLSI circuits. As a result, improving these circuits' efficiency can improve larger computing blocks' performance [10].

A number of ternary logic computational circuits are given in [11-22]. A multiplexer-based ternary half adder and a 1-trit multiplier are presented in [14]. A non-simplified ternary half adder is suggested based on 3:1 multiplexers in [15]. Ternary half adder and 1-trit multiplier designs based on decoders, inverter, and NAND gates are described in [16]. A general method for designing ternary functions is investigated in [17], and a half adder circuit is designed based on the concept. A 1-trit multiplier is also proposed using two types of 2:1 multiplexers in [18]. In [19] and [20], ternary gates have been designed without using the $V_{DD}/2$ voltage source, and subsequently, Ternary Half Adder (THA) and Ternary 1-trit Multiplier (TMUL) circuits are implemented using the input decoding method and subsequent binary and ternary gates. Recently, In [21], a dynamic ternary half adder (DTHA) using a dynamic encoder and two power supplies has been designed. In [22], using unary operators, a THA with a small number of transistors is designed. Ternary adder and multiplier designs without using decoders or encoders have recently been analyzed in [23].

In previous research, a dynamic ternary full adder was designed using CNFETs. The results were published in [11]. Since large arithmetic modules such as multipliers ultimately require smaller units such as half adders in the partial product reduction process, the design of half adder and 1-trit multiplier were targeted in this paper. Eventually, using all the cells, including the full adder, half adder, and 1-trit multiplier, larger arithmetic blocks such as multiple trit multipliers can be designed. If full adders are to be used instead of half adders at all instances, that can lead to unnecessary power dissipation [12,13]. A multi-trit multiplier design analysis is reviewed in [12]. As an example, Figure (1) shows how to multiply two 4-trit numbers. The multiplier of Figure (1) uses 16 TMULs, 15 THAs, and 18 Ternary Full Adders (TFA).

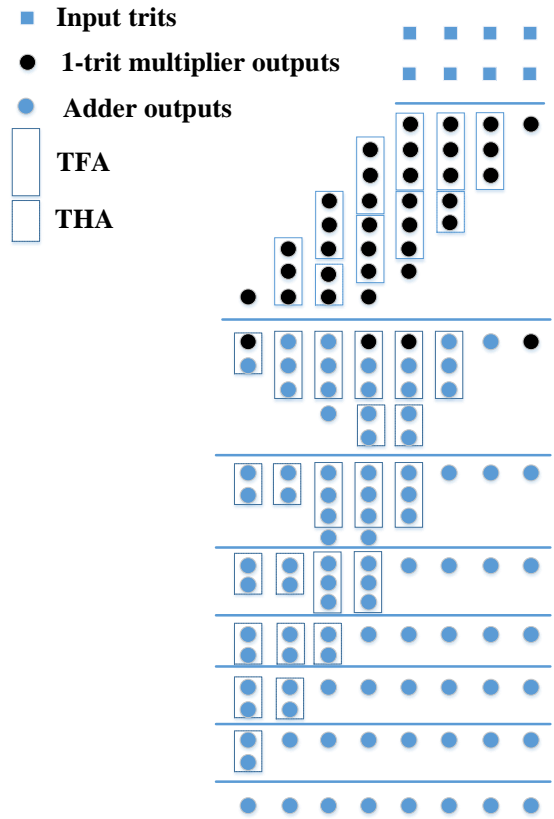

Figure 1. A 4×4 trit multiplier

Every Boolean function can be implemented using pass transistor logic. In this method, pass transistors (p-CNFET and n-CNFET) and transmission gates are used to design the circuit [24]. The method presented here also uses dynamic logic [25] to reduce dynamic power dissipation. The dynamic logic reduces the number of transistors by removing one of the pull-up or pull-down networks, resulting in a significant reduction in the unit's input capacitance. With the dynamic logic, the processing is divided into two phases, pre-charging (or pre-discharging) and evaluation. In the first phase, independent of the inputs, the output node is fully charged or discharged. In the evaluation phase, the output value becomes valid. Depending on the combination of inputs, the output node either retains its previous value or is discharged (or charged) by the rest of the network. In the paper, it is shown that the combination of the pass transistor and dynamic approach can be an effective method for improving the performance of ternary half adder and 1-trit multiplier logic gates.

The rest of this paper is organized as follows: Section 2 provides a brief overview of carbon nanotube transistors. Section 3 discusses the ternary half adder and 1-trit multiplier circuits in the literature. The proposed method is described in Section 4. In Section 5, the proposed method is evaluated compared to other methods, and in the end, the article is summarized in Section 6.

## 2. Carbon Nanotube Field-Effect Transistor

A carbon nanotube is an allotrope of carbon with a cylindrical shape and can be considered a tubular graphene sheet. Carbon nanotubes have unique electronic properties that are determined by how the sheet is tubed [26].

Depending on the number of tubes, carbon nanotubes are divided into single-walled carbon nanotubes (SWCNTs) and multi-walled carbon nanotubes (MWCNTs). Single-layer carbon nanotubes consist of a single-cylinder, and multilayer carbon nanotubes consist of more than one cylinder stacked inside each other. One-dimensional ballistic electron transfer is one of the most important advantages of single-walled carbon nanotubes [27]. The nanotube is formed by taking one atom of the sheet and rolling it onto the atom that is located at $n_1 \times \vec{a_1} + n_2 \times \vec{a_2}$ away from the original atom; where $\vec{a_1}$ and $\vec{a_2}$ are unit vectors [26]. Typically, the chiral vector of carbon nanotubes determines its electronic characteristics based on the $n_1$ and $n_2$ indices. The direction of the chiral vector elements and the direction in which the graphene sheet is tubed determines whether the single-walled carbon nanotube is metallic or semiconductor. If $n_1$-$n_2$ is a multiple of 3, the SWCNT will be metallic; otherwise, it will be a semiconductor [27]. Semiconductor carbon nanotubes can be used as a channel for carbon nanotube transistors [28, 29].

Single-walled carbon nanotubes are divided into three types based on the chiral vector, characterized by indices ($n_1$, $n_2$): armchair, chiral, and zigzag. If $n_2 = 0$ or $n_1 = 0$, the nanotube will be zigzag, if $n_1 = n_2$, the nanotube will be armchair and if $n_1 \neq n_2 \neq 0$, the nanotube will be chiral [28].

CNFET can be more energy-efficient than MOSFET, which is due to Ballistic performance, negligible surface scattering, significantly less parasitic effects, and leakage current. In addition, the one-dimensional structure of these transistors reduces the resistance and thus improves energy efficiency [6,26]. The significantly higher on-to-off current ratio and the same mobility of pCNFET and nCNFET transistors are other important features of CNFETs that reduce the design complexity and size of CNFET-based circuits compared to MOSFET-based circuits [29]. The threshold voltage of a CNFET is equal to the voltage required to create a conductor channel between the source and drain and turn on the transistor. As can be seen in Equation (1), one of the most important features of CNFET is that its threshold voltage can be determined according to the diameter of the nanotubes located under the transistor's gate [29].

$$V_t \cong \frac{E_g}{2e} = \frac{\sqrt{3}}{3}\frac{aE_\pi}{D_{CNT}} \cong \frac{0.43}{D_{CNT}} \quad (1)$$

In Equation (1) $a$ ($\cong 0.249 nm$) is the carbon to carbon atom distance, $E\pi$ ($\cong 3.033 eV$) is the carbon π–π bond energy in the tight bonding model, e is the unit electron charge, and $D_{CNT}$ is the diameter of the nanotubes in nanometers which is calculated with the use of Equation (2).

$$D_{CNT} = \frac{a \times \sqrt{n_1^2 + n_1^2 + n_1 n_2}}{\pi} \cong 0.0783 \sqrt{n_1^2 + n_2^2 + n_1 n_2} \quad (2)$$

From the previous relations, it can be concluded that changing the chiral vector indices changes the difference in energy bands of carbon nanotubes and finally changes the CNFET threshold voltage. The ability to determine the threshold voltage by changing $n_1$ and $n_2$ is a prominent feature of this type of transistor that distinguishes it from MOSFET. This feature makes CNFET suitable for designing voltage-mode MVL circuits based on multiple threshold voltages. Many effective methods for developing carbon nanotubes with specified chirality and obtaining the desired threshold voltage for multi-tube CNFETs have been proposed in previous research. Although CNFET-based designs have fewer carbon nanotubes of different diameters, they can reduce the complexity of fabricating CNFET-based integrated circuits [28-31].

So far, three types of CNFET have been introduced. The first type of carbon nanotube transistor is called the Schottky barrier CNFET (SBCNFET), where the source and drain areas are made of metal, and as a result, there is a Schottky barrier at the drain and source terminals. An important point about this type of CNFET is the adaptation of Schottky barriers and gate electrodes, which may cause manufacturing problems. Also, the on-current to off-current ratio and strong bipolar characteristics reduce the use of this type of CNFET in the design of high-performance digital circuits. MOSFET-like transistors are the second type of CNFET. In this transistor type, carbon nanotubes are highly doped, so similar MOSFET features are provided in this transistor type.

Besides, the appropriate on-to-off-current ratio makes this type of CNFET suitable for designing high-performance and energy-efficient circuits. The third type of transistor is called a tunneling CNFET (TCNFET), with excellent off-state characteristics but low on-current, which is suitable for very low power applications [28,32].

Considering the advantages and disadvantages of different types of CNFETs that were mentioned above, in the proposed design, the MOSFET-like type has been used to achieve the goal of reducing energy consumption.

## 3. Review of ternary logic design and previous works

Processing, storing, and transmitting large amounts of information in digital signal processing are the main challenges of binary systems. Ternary logic can be used as an alternative method to solve these problems. The advantages of ternary circuits and systems such as faster computational operations, higher storage density, less complexity, fewer connections, and easier testing make it suitable for energy-efficient design [33].

It should be noted that each digit in base 3 is called a Trit. In base 3, a set of balanced or unbalanced digits can be used. In a balanced set, each digit can be equal to '-1', '0', or '1', but in an unbalanced set, each digit can be equal to '0', '1', or '2' [34]. In this paper, only the unbalanced ternary set will be used, and in circuits designed in voltage mode, voltage levels of 0, $V_{DD}/2$, and $V_{DD}$ will be assigned to these values, respectively.

Table (1) shows the maximum presentable unsigned integer for ternary and binary logics at different numbers of digits to show the presented ternary methods' superiority compared to their binary counterparts. According to this table, the number of digits that can be represented with 64 trits is more than $10^{11}$ times the number of digits that can be produced with 64 bits of a binary value.

Table 1. Comparison of the maximum unsigned integer representable by binary and ternary logic gates

| Logic<br>Number of Digits | Binary | Ternary |
|---|---|---|
| 1 | 1 | 2 |
| 2 | 3 | 8 |
| 4 | 15 | 80 |
| 8 | 255 | 6560 |
| 16 | 65535 | 43046720 |
| 32 | 4294967295 | 1.85302018×10$^{15}$ |
| 64 | 1.84467440×10$^{19}$ | 3.43368382×10$^{30}$ |

The ideal noise margin of ternary and binary logics at different supply voltages can be seen in Table (2). The figures in Table (2) are calculated using Equation (3), where $NM$ is the noise margin, and $r$ is the radix [9]. Based on this equation, ternary logic's ideal noise margin is half of the binary logic's noise margin for a constant supply voltage.

$$NM = \frac{V_{DD}}{2r - 2} \quad (3)$$

Table 2. Comparison of the ideal noise margin of binary and ternary logics at different supply voltages

| Logic<br>Supply Voltage (V) | Binary | Ternary |
|---|---|---|
| 3 | 1.5 | 0.75 |
| 1.8 | 0.9 | 0.45 |
| 1.2 | 0.6 | 0.3 |
| 0.9 | 0.45 | 0.225 |
| 0.75 | 0.375 | 0.1875 |

One of the most popular methods for implementing a ternary half adder and a ternary 1-trit multiplier circuit is to use multiplexers. For example, the designs presented in [14], [15], and [18] are reviewed here. In [14], the input signals are decoded; the use of each decoder adds 10 transistors. Also, some multiplexers' inputs are connected to the input signals, prolonging the critical path and increasing the delay in a multi-stage design. In [15], first, a ternary 3:1 multiplexer with 15 transistors is proposed, and then a half adder circuit is designed using 6 multiplexers. This design method simplifies the design process but increases the number of transistors. In practice, some of the same calculations are performed three times, and this circuit can be simplified; also, in areas where the logic value passing through the multiplexer is known, there is no need to use a transmission gate, and only one pass transistor is sufficient. In [18], first, the internal structures of two types of 2:1 MUXs are proposed, and then the final circuit is designed using 5 multiplexers. This circuit also can be simplified and can have a long delay in connecting successive stages.

In [16], first, a ternary inverter, NAND, and decoder is proposed, and then each of the half adder inputs is decoded, and the outputs are entered into an array of NAND gates and inverters. Disadvantages of this method include the presence of many gates, which increases the power consumption due to increasing the number of transistors, increasing the activity factor of intermediate nodes, and increasing the delay due to the large capacitances.

In [17], the semiconductor circuit is designed using the general method of simplifying functions with the Karnaugh map. The advantage of this method is its ability to be applied to any desired ternary function, and its main disadvantage is in designing three sub-circuits and connecting their outputs to two encoders, which causes several nodes with high activity factors and thus increases power consumption.

By comparing the previous works, it can be seen that the use of pass transistor logic reduces the number of transistors compared to structures to complementary pull-up network (PUN) and pull-down network (PDN). It should be noted that other designs that have been proposed in recent years will not be compared with the proposed design due to the differences in the design methods and the design objectives; Such as designs that use capacitive voltage division at the input or designs that do not use a $V_{DD}/2$ power supply and determine the final output by voltage division or by causing a voltage drop. In circuits that use only one voltage source, voltage dividers are used to generate $V_{DD}/2$ voltage for logic '1' (by resistors or ON transistors that provide a path from the power supply to ground or diode voltage drops). These circuits draw high static currents, resulting in significant power dissipation, and are thus not suitable for low-power applications. Also, circuits that use capacitive voltage dividers at their input or output have lower drive power, and larger capacitors must be used to compensate for this, which increases the area. In this research, the goal is to design low-power circuits that do not occupy a large area and have suitable drive power.

In order to connect successive stages of dynamic ternary circuits without creating problems such as leakage and clock feedthrough, methods such as using ternary latches and two-phase clocks have been proposed in [11].

## 4. Proposed Circuits

In the proposed circuits, The gates shown in Figure (2) are used as inverters. The truth table of these inverters can be seen in Table (3). Also, the chiral vectors used in the proposed design, along with the diameter of the nanotubes and the equivalent threshold voltages, are shown in Table (4). The threshold voltage of each CNFET can determine at which logic level the CNFET is turned on.

Since the ternary values '0', '1', and '2' are represented with voltage levels of 0, $V_{DD}/2$, and $V_{DD}$, respectively, the logic levels that can activate each specific CNFET can be described as shown in Table (5).

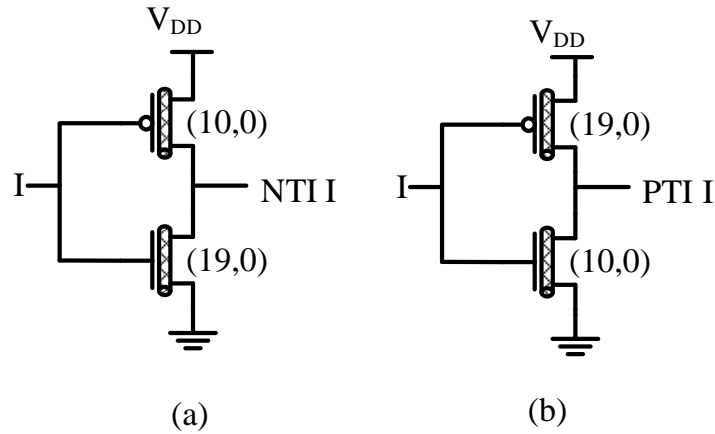

Figure 2. Structure of (a) Negative Ternary Inverter (NTI) and (b) Positive Ternary Inverter (PTI) used in the proposed circuits

Table 3. Negative Ternary Inverter (NTI), Positive Ternary Inverter (PTI), and Standard Ternary Buffer (STB) 's truth tables

| I | NTI I | PTI I | STB I |
|---|---|---|---|
| '0' | '2' | '2' | '0' |
| '1' | '0' | '2' | '1' |
| '2' | '0' | '0' | '2' |

Table 4. Chiral vector, diameter, and the threshold voltage of transistors used in the proposed designs

| Chiral Vector | Diameter | Threshold Voltage |
|---|---|---|
| (10,0) | 0.783 nm | 0.549 V |
| (19,0) | 1.488 nm | 0.289 V |
| (28,0) | 2.192 nm | 0.196 V |

Table 5. On or off status of transistors with different threshold voltages for different inputs

| Chirality | Transistor type/Input | 0 | $V_{DD}/2$ | $V_{DD}$ |
|---|---|---|---|---|
| (10,0) | N-CNFET | OFF | OFF | ON |
|  | P-CNFET | ON | OFF | OFF |
| (19,0) | N-CNFET | OFF | ON | ON |
|  | P-CNFET | ON | ON | OFF |
| (28,0) | N-CNFET | OFF | ON | ON |
|  | P-CNFET | ON | ON | OFF |

The CNFETs are arranged in series such that the appropriate combination of inputs connects the output to the desired voltage rail of 0, $V_{DD}/2$, or $V_{DD}$. In a non-dynamic design, the CNFET network should introduce paths from the output to all the 0, $V_{DD}/2$, and $V_{DD}$ supply rails. However, with the proposed dynamic design, one of these paths can be omitted by initially pre-charging/per-discharging the output to one of the supply rails in the pre-charge phase. The pre-charge voltage is chosen such that the node activity is reduced. The proposed circuits are optimized regarding power dissipation by pre-charging the output node to the voltage level that occurs the most. With this approach, the switching activity of the output node can be reduced.

In sections 4.1 and 4.2, the proposed DTHA and DTMUL (Dynamic Ternary Multiplier) will be described, respectively.

**4.1 Proposed Dynamic Ternary Half Adder**

This section describes the proposed ternary half adder based on dynamic logic. In the proposed design, the pass transistor circuit is initially designed according to the ternary half adder truth table shown in Table (6), and subsequently, the dynamic logic properties are added to the resulting design. Table (6) also shows the activated path that eventually determines the state of the output values. First, the circuits' two output nodes are pre-charged and pre-discharged to $V_{DD}/2$ and 0V by transistors C1 and C20, respectively. As a result, all paths consisting of the transistors that connect the first and second output to logic values of '1' and '0', respectively, will be removed. As a result, the parasitic capacitor of the output node and some intermediate nodes is reduced. On the other hand, in 9 possible input combinations, in Sum output, each of the outputs values '0', '1', and '2' are generated 3 times; As a result, the output node's activity will be constant if pre-charged to any of the three logic levels; However, if the output is pre-discharged to '0' or pre-charged to '2', in each output change, the magnitude of the voltage change will be more significant. Nevertheless, in the Carry output logic, the value '2' is never generated, and in two-thirds of the possible cases, the output will be '0'. The final circuit related to the ternary half adder's Sum output can be seen in Figure (3), and the circuit related to the Carry output can be seen in Figure (4). The dashed area is actually a 3:1 MUX. A dashed line with arrows also indicates the critical path. In the circuits, any transistor that its chiral vector is not written next to; has a chirality vector of (19,0).

Table 6. The ternary half adder's truth table

| ON Path | A | B | Carry | Sum |
|---|---|---|---|---|
| C2, C10, C20 | '0' | '0' | '0' | '0' |
| C1, C20 | '0' | '1' | '0' | '1' |
| C3, C11, C20 | '0' | '2' | '0' | '2' |
| C1, C20 | '1' | '0' | '0' | '1' |
| C5, C7, C12, C13, C20 | '1' | '1' | '0' | '2' |
| C4, C6, C14, C21, C22, C24 | '1' | '2' | '1' | '0' |
| C9, C15, C20 | '2' | '0' | '0' | '2' |
| C8, C16, C17, C23, C25, C26 | '2' | '1' | '1' | '0' |
| C1, C23, C27 | '2' | '2' | '1' | '1' |

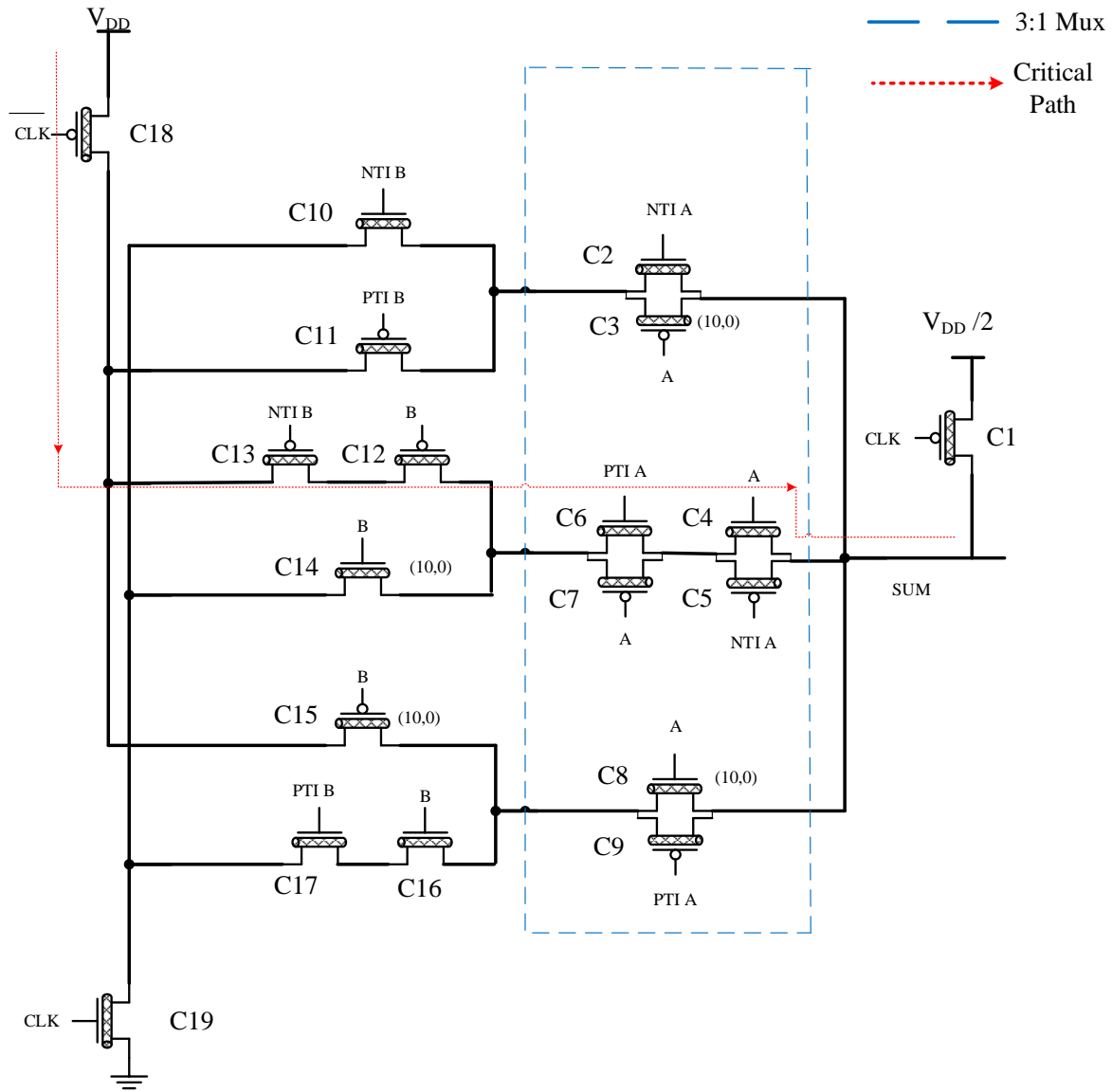

Figure 3. Proposed ternary half adder's sum generator circuit

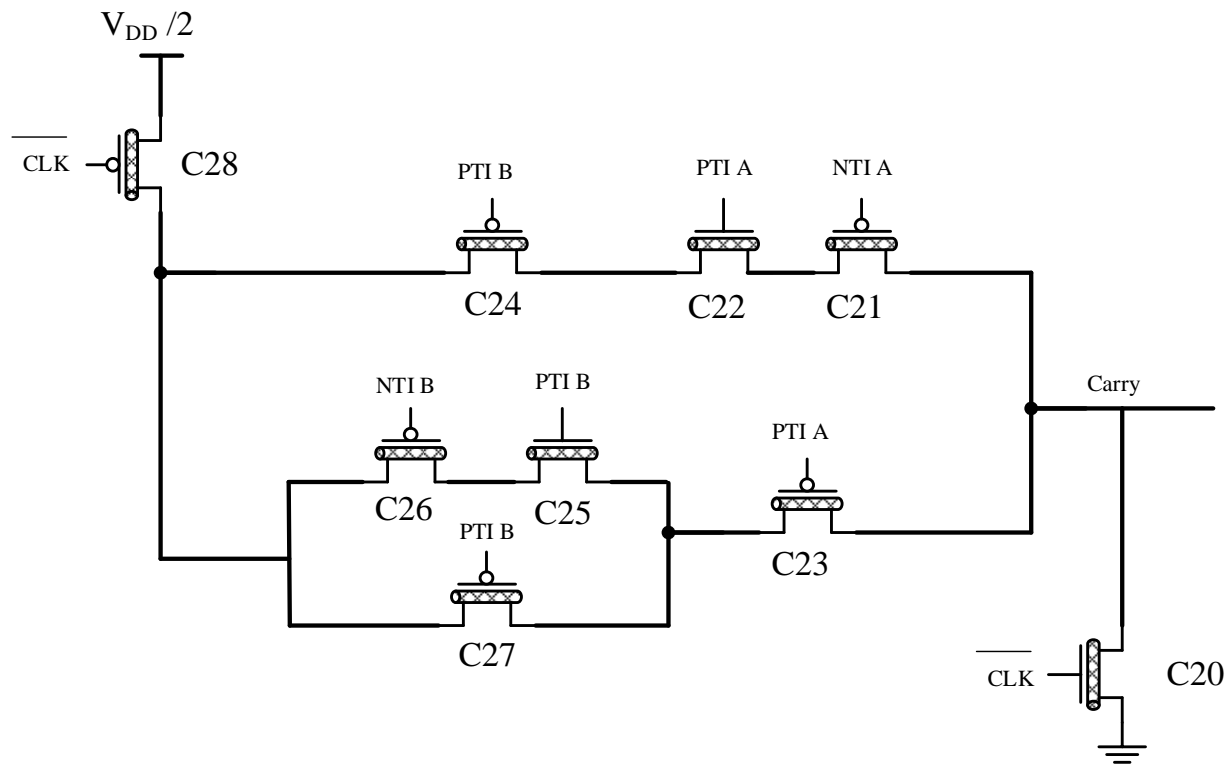

Figure 4. Proposed ternary half adder's carry generator circuit

### 4.2 Proposed Dynamic Ternary 1-trit Multiplier

This section describes the proposed ternary 1-trit multiplier based on dynamic logic. The 1-trit multiplier's truth table is shown in Table (7). The activated path that eventually determines the state of the output trits is also shown in Table (7). Similar to the half adder, the circuit based on the pass transistor logic is designed, and then the properties related to dynamic logic are added. During operation, in the pre-charge phase, the circuits' output nodes are pre-discharged to 0V using transistors with (19,0) chiral vectors. As a result, all paths consisting of the transistors that connect the output to the logic value '0' will be deleted, and the parasitic capacitor of the output node and some intermediate nodes is reduced. On the other hand, in 9 possible input combinations, output '0' is generated in 5 instances in Product output, and output values '1' and '2' are each generated in only 2 cases. In the Carry circuit, the output is not '0' for just one of 9 possible states; Thus, the output node's activity factor will be less if it is pre-discharged to the level of '0'. The final circuit related to the 1-trit multiplier's Product output is shown in Figure (5), and the circuit related to the 1-trit multiplier's Carry output is shown in Figure (6). The critical path is also indicated by a dashed line with arrows. In these figures, any transistor that its chiral vector is not written next to it has a chirality of (28,0). As can be seen in Table (5), the transistors with chiralities of (19,0) and (28,0) have similar on or off states for 0, $V_{DD}/2$, and $V_{DD}$ inputs. In Figures (5) and (6), if these transistors have the same chirality, the circuits will operate correctly however the performance will be degraded. If all CNFETs with (28,0) chirality are substituted with (19,0), the power consumption will be less than the proposed circuit but the latency increases abruptly, and as a result, the PDP and EDP of the circuit gets worse. Subsequently, if all CNFETs with (19,0) chirality are substituted with (28,0), the power and latency and consequently EDP will be slightly higher than the proposed design. Table (8) shows the performance change for the three different described scenarios.

Table 7. The 1-trit multiplier's truth table

| ON Path | A | B | Carry | Product |
|---|---|---|---|---|
| C1, C16 | '0' | '0' | '0' | '0' |
| C1, C16 | '0' | '1' | '0' | '0' |
| C1, C16 | '0' | '2' | '0' | '0' |
| C1, C16 | '1' | '0' | '0' | '0' |
| C4, C5, C6, C7, C16 | '1' | '1' | '0' | '1' |
| C12, C13, C14, C16 | '1' | '2' | '0' | '2' |
| C1, C16 | '2' | '0' | '0' | '0' |
| C9, C10, C11, C16 | '2' | '1' | '0' | '2' |
| C2, C3, C17, C18 | '2' | '2' | '1' | '1' |

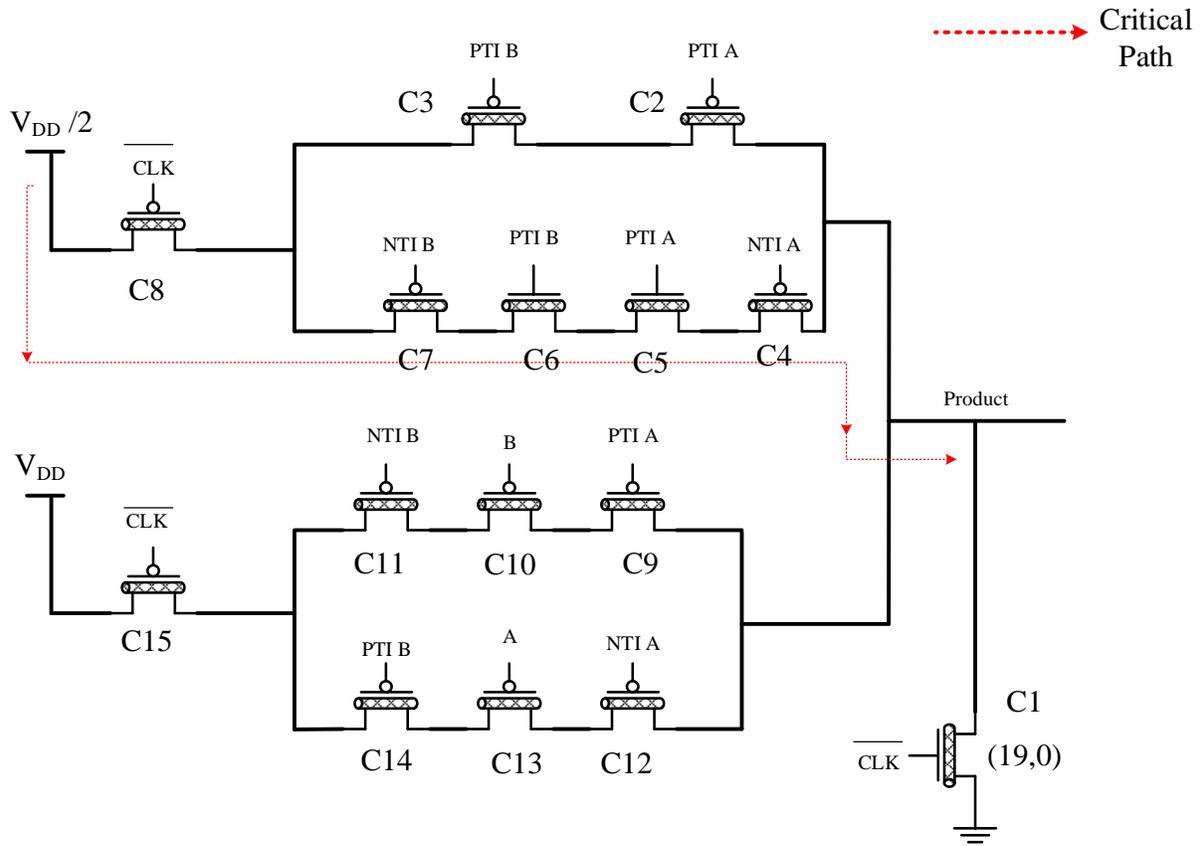

Figure 5. Proposed 1-trit multiplier's product generator circuit

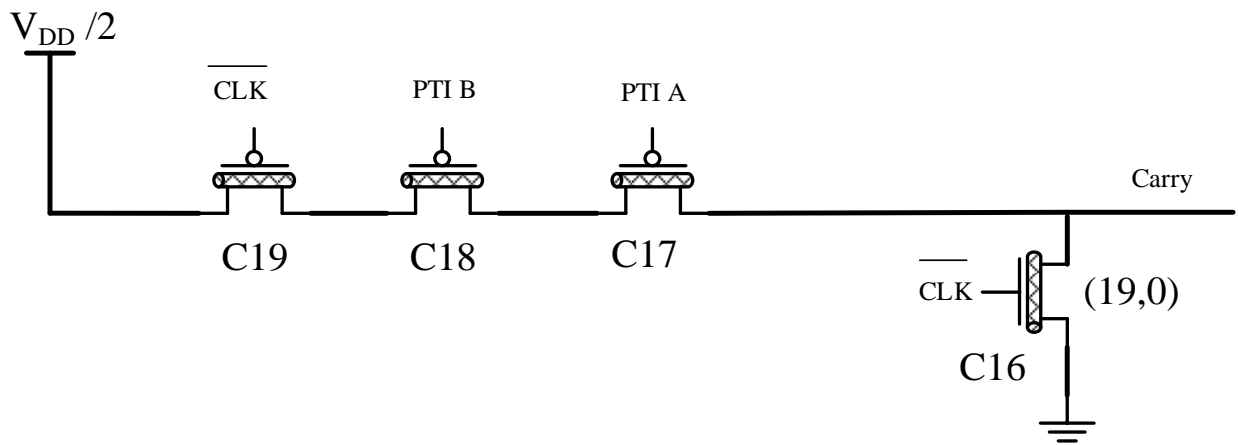

Figure 6. Proposed 1-trit multiplier's carry generator circuit

Table 8. Proposed DTMUL structure's simulation results in 3 different scenarios

| Chiralities | Average Power Consumption (nW) | Maximum Delay (ps) | PDP (aJ) | EDP ($\times 10^{-30}$ J.s) |
|---|---|---|---|---|
| Proposed | 68.22 | **45.41** | **3.09** | **140.71** |
| (28,0) chirality substituted with (19,0) | **60.91** | 175.57 | 10.69 | 1877.4 |
| (19,0) chirality substituted with (28,0) | 68.86 | 45.74 | 3.15 | 144.03 |

## 5. Simulation Results and Comparisons

In this section, the simulation results of the circuits are presented. All designs are simulated by Synopsis HSPICE software using the Stanford 32nm CNFET model [35-37].

Previous designs have been simulated using the values provided for CNFET parameters in the relevant papers and compared with the proposed circuits. All simulations were performed with a supply voltage of 0.9V and a frequency of 1 GHz. To simulate circuits in a realistic condition, a test bench that consists of an STB at each input and four Standard Ternary Inverters (STIs) at each output has been used. Figures (7) and (8) show the transient response of the proposed half adder and 1-trit multiplier circuits to all possible inputs, respectively, which produce correct outputs in all cases.

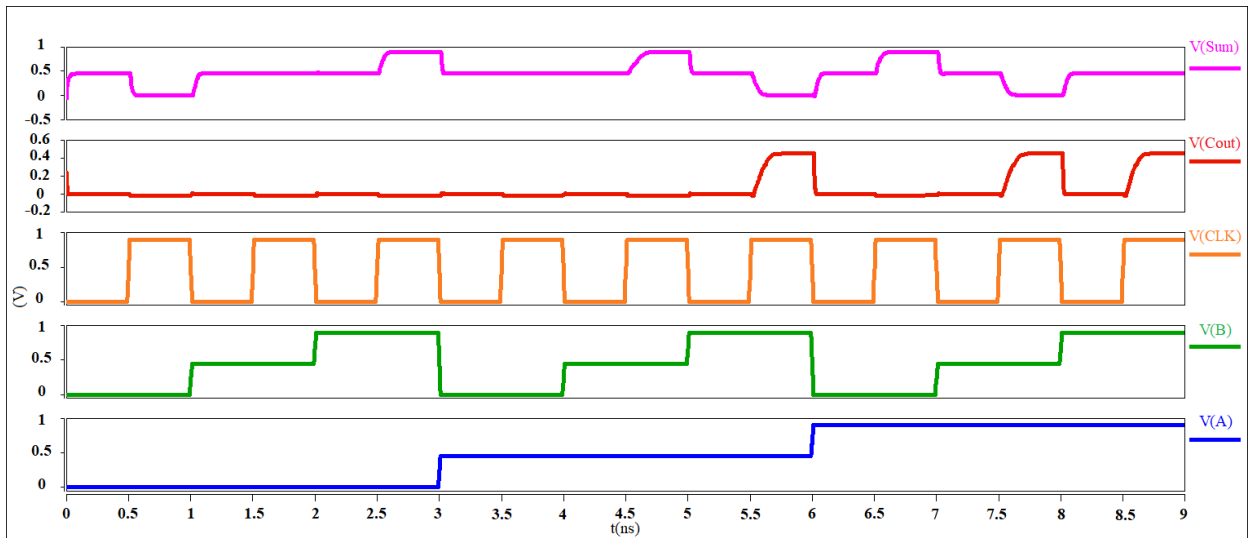

Figure 7. The transient response of the proposed half adder to 9 different input states

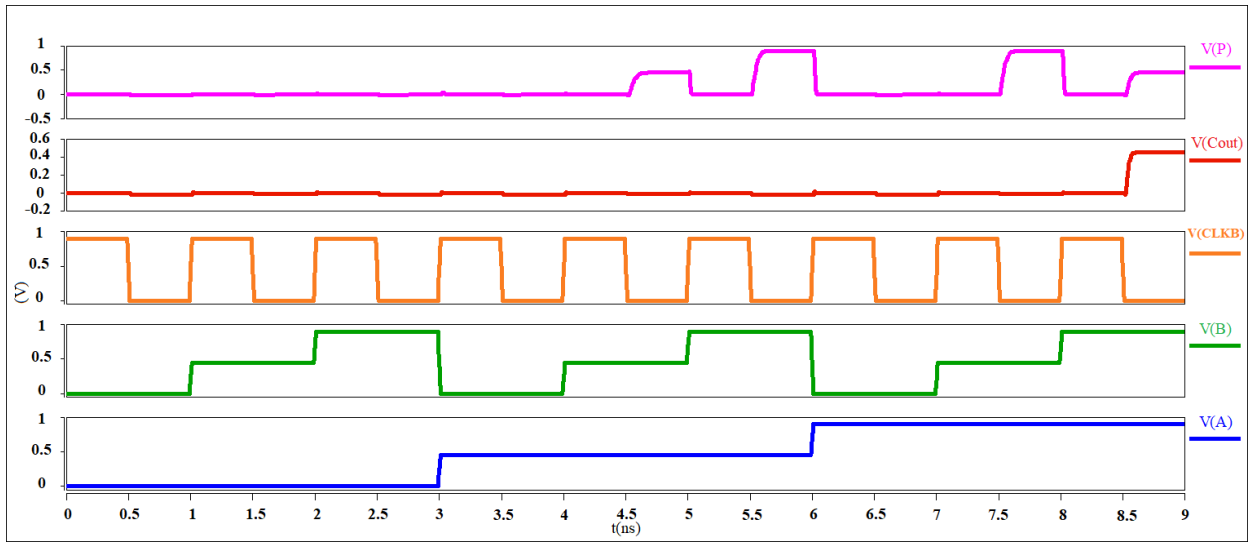

Figure 8. The transient response of the proposed 1-trit multiplier to 9 different input modes

The input change pattern for simulating the two input THA and TMUL gates, includes all 81 possible states of input change. Circuit delay is calculated for all output states from circuit input (after buffers) to circuit output (before FO4). In terms of power consumption, in addition to switching output nodes, the activity of intermediate nodes also affects the power consumption of the circuit. Therefore, in order to accurately evaluate the power consumption, the average power consumption is considered for all cases of change of inputs. To show the efficiency of the proposed circuit and the trade-off between propagation delay and power consumption, PDP and EDP are considered as FOMs, and their values have been reported. The results of the simulations related to the proposed ternary half adder and 1-trit multiplier and their comparison with other designs can be seen in Tables (9) and (10), respectively. Best figures of merit are highlighted in bold.

Table 9. Simulation results of ternary half adders

| THA Design | Power (nW) | Static Power (nW) | Delay (ps) | PDP (aJ) | EDP ($\times 10^{-30} J.s$) | No. of transistors | No. of chiralities |
|---|---|---|---|---|---|---|---|
| Proposed | **84.84** | **1.99** | 73.25 | **6.21** | **455.33** | 36 | 2 |
| [14] | 133.70 | 16.88 | 84.20 | 11.25 | 947.96 | 52 | 2 |
| [15] | 237.53 | 22.54 | 138.25 | 32.83 | 4540.1 | 90 | 2 |
| [16] | 405.81 | 208.72 | 85.45 | 34.67 | 2963.3 | 85 | 3 |
| [17] | 182.75 | 3.65 | 321.73 | 58.79 | 18917 | 50 | 4 |
| [19] | 1142.0 | 294.98 | **54.76** | 62.54 | 3425.1 | 112 | 3 |
| [20] | 1338.6 | 489.74 | 76.36 | 102.22 | 7806.2 | 124 | 3 |
| [22] | 138.38 | 85.54 | 113.73 | 15.74 | 1790.0 | **34** | 2 |

The proposed half adder's power consumption is 37%, 64%, 79%, 54%, 39%, 94%, 93% less than [14], [15], [16], [17], [22], [20] and [19] designs, respectively. Propagation delay is also decreased by 13%, 47%, 14%, 77%, 36% and 4% respectively. Also PDP parameter improved by 45%, 81%, 82%, 89%, 61%, 94%, and 90% and EDP parameter improved by 52%, 90%, 85%, 98%, 75%, 94% and 87% respectively. The static power of the proposed design is also less than its previous counterparts. The proposed design has 14, 16, 24, 49, 54, 76, and 88 transistors less than [17], [14], [21], [16], [15], [19], and [20] designs, respectively, which are equivalent to 28%, 31%, 40%, 58%, 60%, 68%, and 71% reduction in transistor count, respectively. In [14], [15], [22], and the proposed designs, only CNFETs with two types of threshold voltages have been used, making their fabrication easier.

Table 10. Simulation results of 1-trit multipliers

| TM Design | Power (nW) | Static Power (nW) | Delay (ps) | PDP (aJ) | EDP ($\times 10^{-30} J.s$) | No. of transistors | No. of chiralities |
|---|---|---|---|---|---|---|---|
| Proposed | 68.22 | **0.65** | 45.41 | 3.09 | 140.71 | 25 | 3 |
| [14] | **45.45** | 3.91 | 55.41 | **2.52** | **139.53** | **23** | 3 |
| [16] | 422.55 | 305.1 | 70.71 | 29.88 | 2112.9 | 61 | 3 |
| [18] | 61.92 | 6.41 | 98.08 | 6.07 | 595.73 | 32 | **2** |
| [19] | 681.94 | 111.9 | **44.25** | 30.18 | 1335.3 | 80 | 3 |
| [20] | 773.48 | 168.6 | 78.85 | 60.98 | 4808.8 | 92 | 3 |

The proposed 1-trit multiplier's delay is 18%, 36%, 42%, and 54%, less than that of [14], [16], [20], and [18], respectively. The proposed circuit's average power consumption is about 84%, 90%, and 91% better than [16], [19], and [20] designs, respectively, but it is 10% and 50% higher than [18] and [14]'s circuits, respectively. Also, the proposed design's static power is less than other designs, which can reduce the average power consumption at lower frequencies. Also, the PDP parameter improved by 49%, 90%, 90%, and 95% and the EDP parameter improved by 76%, 93%, 90%, and 97% compared to [18], [16], [19], and [20]'s circuits, respectively. The proposed design has 67, 55, 36, 7 fewer transistors than [20], [19], [16] and [18]'s designs, respectively, which are correspondingly equivalent to 73%, 69%, 59% and 22% reduction.

As shown in Equation (4), the total power dissipation, $P_{Total}$, is dependent on the static power dissipation, $P_{Static}$ and dynamic power dissipation, $P_{Dynamic}$. The transient short circuit power dissipation can also be considered by modifying the equivalent capacitance of the dynamic power dissipation term.

$$P_{Total} = P_{Static} + P_{Dynamic} \quad (4)$$

Equation (5) shows that the static power dissipation is dependent on the average subthreshold, gate, junction, and contention currents, $I_{sub}$, $I_{gate}$, $I_{junct}$ and $I_{contention}$ respectively. The proposed design has a very low static power dissipation due to zero contention currents. Usually, circuits with high contention currents will show higher power dissipation while their delay is smaller due to the lower number of transistors required in the design.

$$P_{Static} = (I_{sub} + I_{gate} + I_{junct} + I_{contention}) \times V_{DD} \quad (5)$$

According to Equation (6), the dynamic power dissipation is dependent on the total number of nodes, *n*, node capacitance effects, $C_i$ and node activity factors, $α_i$ [38,39]. Designs with lower number of transistors, such as those presented in this paper, can help reduce the dynamic power usage by decreasing the capacitance effects; however, with the proposed designs, since the dynamic logic approach is used in the circuit, the activity factor at the output node can be increased compared with conventional design due to the continuous output node pre-charging. Thus depending on the more dominant power dissipation term, different performance factors will be observed. [14] and [18] have better power dissipation because of the lack of a dynamic pre-charging circuit. However, as it is evident from the results, since logic gates with pre-charge structures can reduce the number of transistors in the critical path, the proposed design will show better delay performance than [14] and [18].

$$P_{Dynamic} = \sum_{i=1}^{n} α_i \times f \times C_i \times \Delta V^2 \quad (6)$$

As shown in Equation (7), the delay, $t_{pd}$, is dependent on the number of transistors in the critical path, *m*, the equivalent resistance on the critical path to node *j*, $R_j$, and the capacitive load at each node, $C_j$ [38]. The proposed approach has a lower delay than most designs due to the low transistor count in the critical path; however, designs such as [19] have a smaller delay at the cost of high contention currents and thus higher power dissipation.

$$t_{pd} = \sum_{j=1}^{m} R_j C_j \quad (7)$$

### 5.1 Supply voltage variations

The proposed ternary half adder and other designs are simulated for supply voltages of 0.8V, 0.9V, and 1V, and the results can be seen in Figure (9). Based on the simulation results at all supply voltages, the proposed design has less power consumption, delay, and consequently less PDP and EDP.

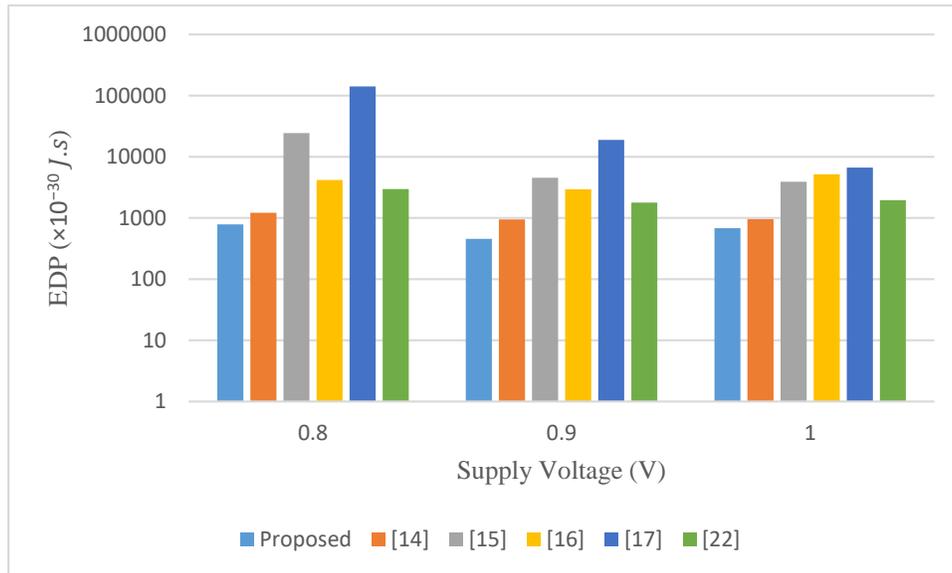

Figure 9. The effect of power supply voltage changes on the simulation results of half adders

The proposed 1-trit multiplier and other designs are also simulated for supply voltages of 0.8V, 0.9V, and 1V, and the results are shown in Figure (10). The noteworthy point is that the proposed circuit's EDP is lower than the others, at 0.8V and 1V supply voltages.

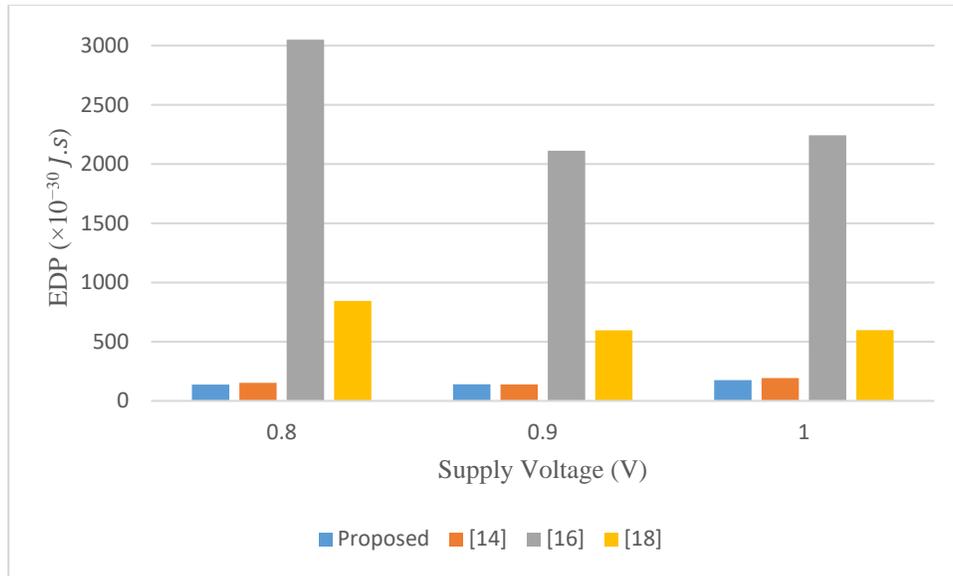

Figure 10. The effect of power supply voltage changes on the simulation results of 1-trit multipliers

## 5.2 Output load variations

Half adder outputs have been connected to 1, 2, and 4 standard ternary inverters, and the respective simulation results can be seen in Table (11). The proposed design maintains its superiority in each parameter by changing the output load.

Table 11. The effect of the output load change on the results of half adder simulations

| Load | THA Design | Power (nW) | Delay (ps) | PDP (aJ) | EDP ($\times 10^{-30} J.s$) |
|---|---|---|---|---|---|
| FO1 | Proposed | **51.79** | **32.41** | **1.68** | 54.38 |
|  | [14] | 112.49 | 70.84 | 7.97 | 564.65 |
|  | [15] | 221.75 | 111.6 | 24.76 | 2764.9 |
|  | [16] | 384.95 | 74.07 | 28.51 | 2111.9 |
|  | [17] | 162.74 | 311.98 | 50.77 | 15839 |
|  | [22] | 121.57 | 49.45 | 6.01 | 297.24 |
| FO2 | Proposed | **63.88** | **46.08** | **2.94** | 135.63 |
|  | [14] | 117.90 | 75.65 | 8.92 | 674.68 |
|  | [15] | 226.71 | 117.79 | 26.70 | 3145.7 |
|  | [16] | 392.91 | 78.37 | 30.79 | 2413.0 |
|  | [17] | 167.82 | 313.4 | 52.59 | 16483 |
|  | [22] | 126.99 | 69.42 | 8.82 | 612.05 |
| FO4 | Proposed | **84.84** | **73.26** | **6.21** | 455.33 |
|  | [14] | 133.70 | 84.20 | 11.26 | 947.96 |
|  | [15] | 237.53 | 138.25 | 32.83 | 4540.1 |
|  | [16] | 405.81 | 85.45 | 34.68 | 2963.3 |
|  | [17] | 182.75 | 321.73 | 58.80 | 18917 |
|  | [22] | 138.38 | 113.73 | 15.74 | 1790.0 |

The 1-trit multiplier outputs have also been connected to 1, 2, and 4 standard ternary inverters, and the corresponding simulation results can be seen in Table (12). The proposed circuit's PDP and EDP parameters are less than others in FO2 and FO1 modes. In the proposed design, the significant increase of PDP and EDP for high fan-outs (FO4) can be attributed to the output node's high signal activity due to the output node's continuous pre-charging. While in [11], for larger computational blocks such as full adders, this does not possess severe performance degradation since the reduction of the number of transistors in the main signal path favors the dynamic power dissipation; however, in smaller designs such as half adders, the dynamic power dissipation at the output node can show some performance degradation at high fan-outs.

Table 12. The effect of the output load change on the results of 1-trit multiplier simulations

| Load | TM Design | Power (nW) | Delay (ps) | PDP (aJ) | EDP ($\times 10^{-30} J.s$) |
|---|---|---|---|---|---|
| FO1 | Proposed | 44.16 | **22.68** | **1.00** | **22.71** |
|  | [14] | **32.18** | 38.26 | 1.23 | 47.10 |
|  | [16] | 408.37 | 60.09 | 24.54 | 1474.6 |
|  | [18] | 46.04 | 55.88 | 2.57 | 143.76 |
| FO2 | Proposed | 51.98 | **30.50** | **1.585** | **48.37** |
|  | [14] | **36.47** | 43.71 | 1.594 | 69.70 |
|  | [16] | 412.09 | 63.19 | 26.04 | 1645.6 |
|  | [18] | 52.01 | 70.31 | 3.66 | 257.07 |
| FO4 | Proposed | 68.22 | **45.42** | 3.10 | 140.71 |
|  | [14] | **45.45** | 55.41 | **2.52** | **139.53** |
|  | [16] | 422.55 | 70.71 | 29.88 | 2112.9 |
|  | [18] | 61.92 | 98.08 | 6.07 | 595.73 |

### 5.3 Temperature variations

To investigate the effect of temperature changes on the behavior of THAs, the operating temperature of the circuits was changed from 0 to 75 °C. The results of changing the temperature are given in Figure (11). As can be seen, with the temperature variations, the proposed design's EDP variations are much less than other designs.

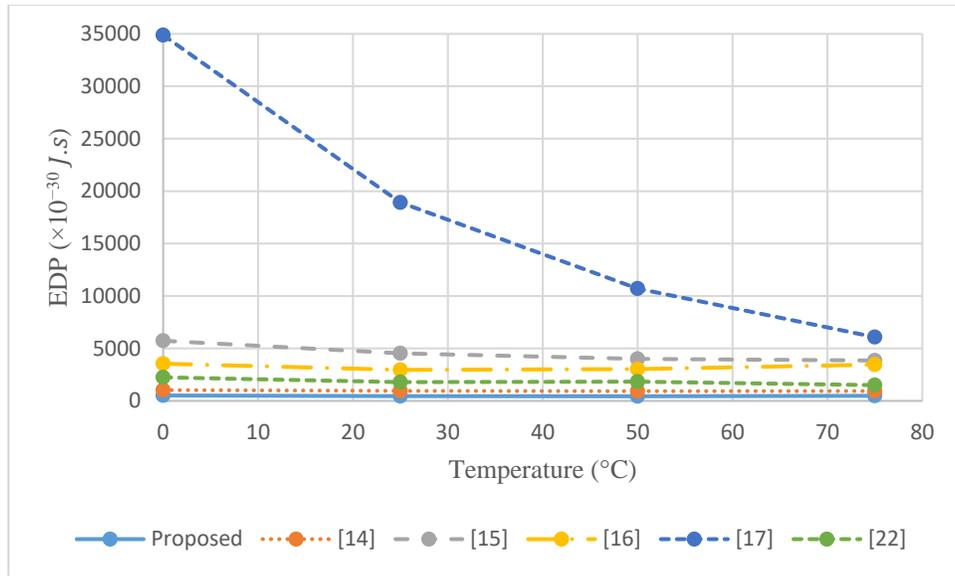

Figure 11. The effect of temperature changes on the simulation results of half adders

The circuit operating temperatures have been changed from 0 to 75 °C to evaluate the 1-trit multiplier's sensitivity to temperature changes. The results are given in Figure (12). As can be seen, the best design in terms of EDP did not change due to temperature variations, except that the proposed method performs better than the others at 75 °C.

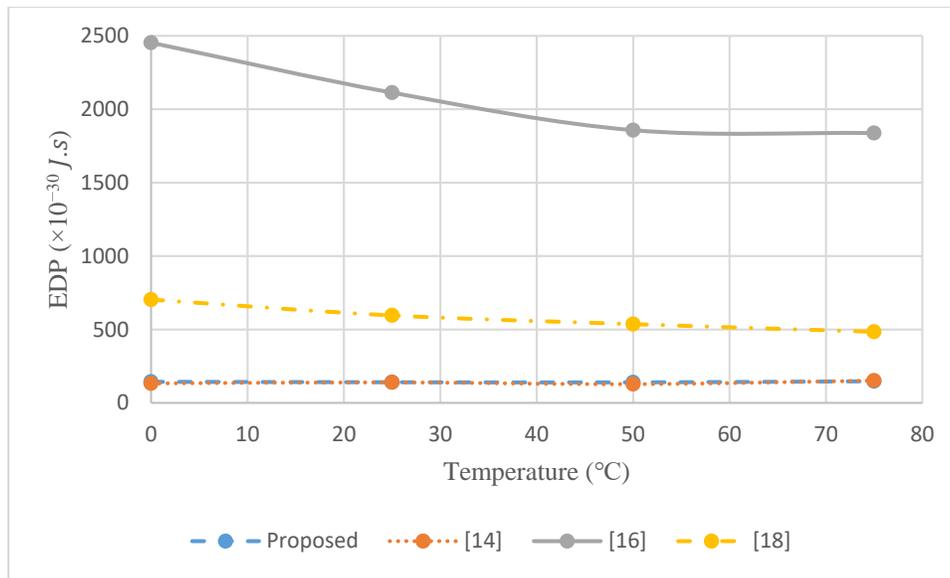

Figure 12. The effect of temperature changes on the simulation results of 1-trit multipliers

## 5.4 Process variations

Process variations in nanoscale devices and circuits have become significant. These parametric changes can have adverse effects on the circuit's speed, power, and reliability [9]. For this reason, to evaluate the robustness of the proposed circuits to process variations, Monte Carlo simulations with 100 iterations are performed. The nanotube diameter, oxide thickness, and nanotube density are parameters that have been changed. It should be noted that to change the diameter, the number $n_1$ of the chiral vector was changed, and to change the density, the number of nanotubes of each transistor and the distance between their centers (Pitch) was changed simultaneously. The parameters mentioned above have been changed under the Gaussian distribution with a variation of 10% and ±3σ. Sigma/mean, which is normalized standard deviation, is a good metric for evaluating and comparing the change rate of circuit parameters due to process variations. Thus it helps evaluate and compare their robustness to process variations [40,41].

The results of these simulations can be seen in Figure (13) and Figure (14). In TMUL cases, the ratio of the variance of changes to the parameter's average for the proposed design was lower than the [14] and [18]'s circuits, which have the closest PDP to the proposed circuit. However, the proposed THA had more PDP variations than [14] and [22].

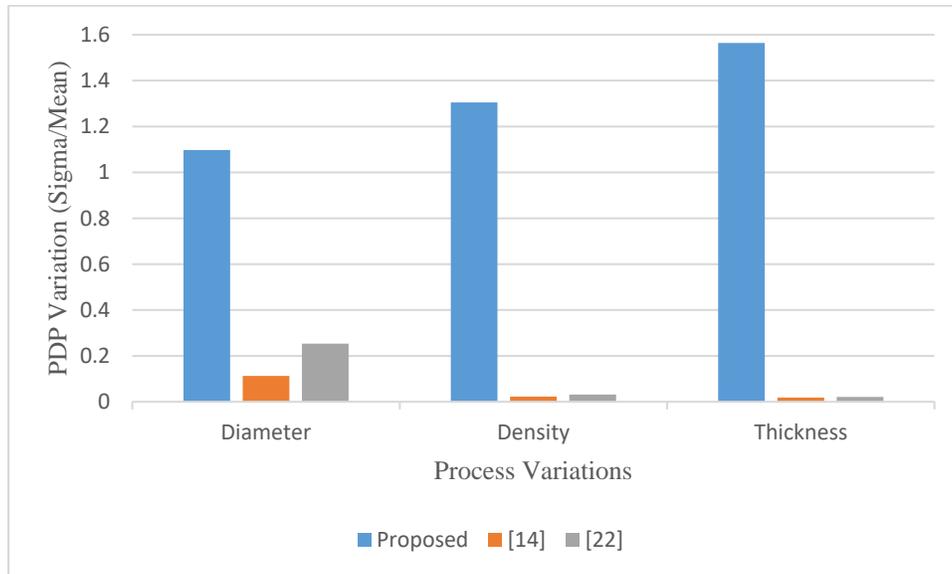

Figure 13. The effect of process variations on the PDP of THAs in terms of the ratio of sigma to mean

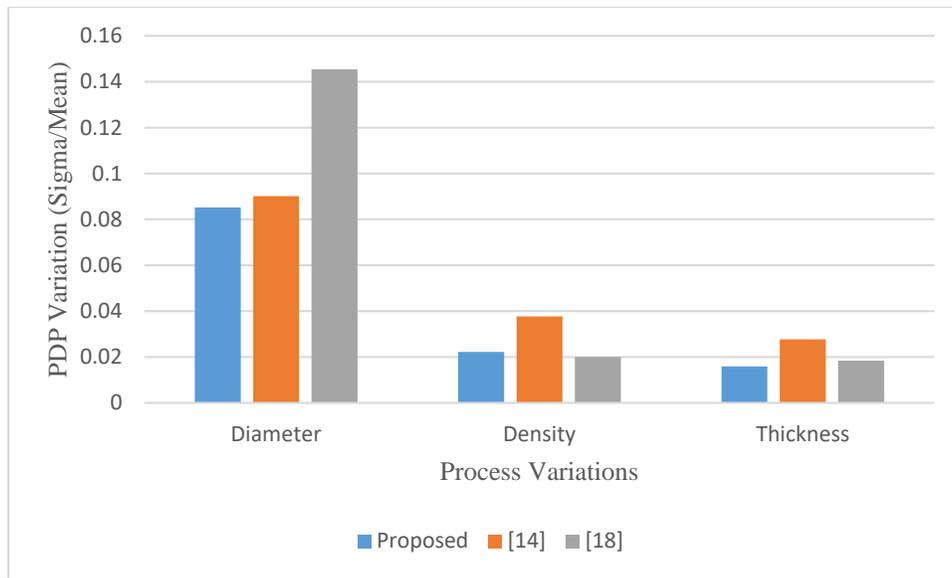

Figure 14. The effect of process variations on the PDP of TMULs in terms of the ratio of sigma to mean

## 6. Conclusion

In this paper, a dynamic ternary half adder and a dynamic ternary 1-trit multiplier were designed and proposed using carbon nanotube transistors. In the proposed circuits, in addition to dynamic logic, pass transistor design approach was also used. The proposed circuits and other similar designs in the literature are simulated, and the simulation results show less propagation delay, power consumption, number of transistors, PDP, and EDP of the proposed half adder compared to other designs. The proposed 1-trit multiplier also has less latency and is better than other designs under small loads in terms of PDP and EDP. The proposed 1-trit multiplier circuit is more robust to process variations compared to designs with the closest PDP.

of Integrated Circuits and Systems 25 (11) (2006) 2427-2436, DOI:

10.1109/TCAD.2006.873886.

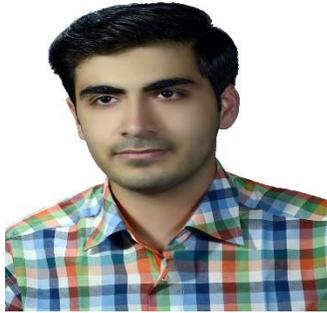

**Farzin Mahboob Sardroudi** received his B.S. degree in Electrical Engineering and Electronics from Shahid Beheshti University, Tehran, Iran, in 2018. He is currently an M.Sc. student in Digital Electronics at the University of Isfahan, Isfahan, Iran. His research interests include digital Integrated Circuit design, low-power VLSI design, Multi-Valued Logic, Carbon Nanotube Field-Effect Transistors, and Quantum-Dot Cellular Automata.

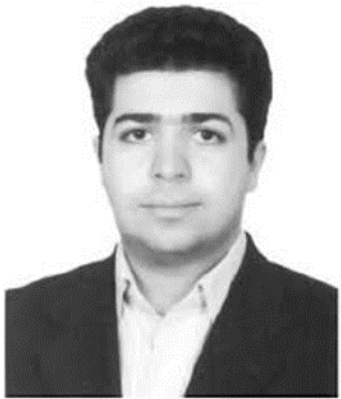

**Mehdi Habibi** was born in Fort Worth, TX, USA, in 1981. He received a B.S., M.Sc., and Ph.D. degrees in electrical engineering from the Isfahan University of Technology, Isfahan, Iran, in 2003, 2005, and 2010, respectively. He is currently an Associate Professor with the Department of Electrical Engineering, University of Isfahan. His current research interests include CMOS sensors and low-power circuit design.

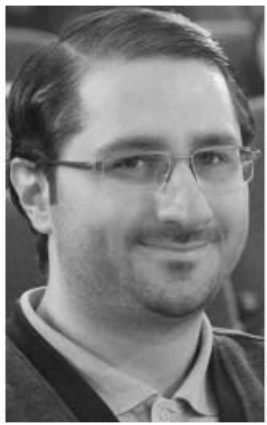

**Mohammad Hossein Moaiyeri** received a Ph.D. degree in computer architecture from Shahid Beheshti University, Tehran, Iran, in 2012. He is currently an Assistant Professor in the Faculty of Electrical Engineering, Shahid Beheshti University, and a Senior Member of IEEE. His research interests include VLSI design for beyond-CMOS emerging nanodevices, low-power VLSI design, and mixed-signal integrated circuits design.